\documentstyle[sprocl]{article}
\input{epsf}
\def\Journal#1#2#3#4{{#1} {\bf #2}, #3 (#4)}
\def\NPB{{\em Nucl. Phys.} B}
\def\PLB{{\em Phys. Lett.}  B}
\def\PRL{\em Phys. Rev. Lett.}
\def\PRD{{\em Phys. Rev.} D}
\def\ZPC{{\em Z. Phys.} C}
\def\be{\begin{equation}}
\def\ee{\end{equation}}
\def\bea{\begin{eqnarray}}
\def\eea{\end{eqnarray}}

\begin{document}
\begin{flushright}
HD-THEP-97-26\\
hep-ph/9706384
\end{flushright}

\title{Hadronic $B$-Decays\footnote{Plenary talk
at the conference on {\it $B$-Physics and CP-Violation}, Honolulu,
Hawaii, March 1997\\
Research supported in part by the Bundesministerium f\"ur
Bildung, Wissenschaft, Forschung und Technologie,
Bonn, Germany.
}}
\author{Berthold Stech}
\address{Institut f\"ur Theoretische Physik, Universit\"at Heidelberg,\\
Philosophenweg 16, D-69120 Heidelberg}
\maketitle\abstracts{
Non-leptonic two-body decays are discussed on
the basis of a generalized factorization approach.
It is shown that a satisfactory description of numerous
decay processes can be given using the same two parameters
$a_1^{eff}$ and $a_2^{eff}$.
Although in general process-dependent, these parameters
are not expected to change markedly. In fact, within error
limits, there is no evidence for a process dependence in
energetic $B$-decays. The success of factorization allows
the determination of decay constants from non-leptonic decays.
For the $D_s$ meson one obtains $f_{D_s}=(234\pm25)$ MeV, for
the $D^*_s$ meson $f_{D_s^*}=(271\pm33)$ MeV. The ratio
$a_2^{eff}/a_1^{eff}$ is positive in $B$-decays and negative
in $D$-decays corresponding to constructive and destructive
interference in $B^-$ and $D^+$ decays, respectively.
Qualitatively, this can be understood considering the different
scales or $\alpha_s$-values governing the interaction
among the outgoing quarks. The running of $\alpha_s$ is also
the cause of the observed strong increase of the amplitude of
lowest isospin when going to low energy transitions.}

\section{Introduction}
The aim of the study of weak decays of hadrons is two-fold:
These decays offer the most direct way to determine the
elements of the Cabibbo-Kobayashi-Maskawa (CKM) matrix
and to explore the physics of CP-violation. At the same
time they are most suitable for the study of strong interaction
physics related to the confinement of quarks and gluons
within hadrons. Both tasks complement each other: an understanding
of the connection between quark and hadron properties is a necessary
prerequisite for a precise determination of the CKM matrix
elements and CP-violating phases.

In non-leptonic decays striking effects have been observed such
as the huge $|\Delta\vec I|=1/2$ enhancement in strange particle
decays, the unexpected liftetime difference of $D^0$ and $D^+$
mesons and the surprising interference pattern in exclusive charged
meson decays. These phenomena led to many speculations and it
was a great challenge to find the correct explanation. Today
we know that the confining colour forces among the quarks are
the decisive factor. By exploring the consequences of the 
QCD-modified effective weak Hamiltonian a semi-quantitative
understanding has been achieved. Still, there is not yet a strict
theoretical approach possible. Matrix elements of local four-quark
operators are hard to deal with. We still have to learn
by confronting models with experiments. Fortunately, thanks
to experimental groups, a huge amount of data is already
available.

In this talk I will concentrate on exclusive non-leptonic
decays of $B$-mesons. For the data I will heavily rely
on the recent detailed report by Browder, Honscheid and Pedrini
\cite{1}. Only two-body decays will be treated.

\section{The effective Hamiltonian}
QCD modifies the simple W-exchange diagram (Fig. 1). Hard-gluon
corrections can be accounted for by renormalization group
techniques. One obtains an effective Hamiltonian incorporating
gluon exchange processes down to a scale of the order of the
heavy quark mass \cite{2}
\begin{eqnarray}\label{1}
&&H_{eff}=\frac{G}{\sqrt2}V_{cb}(c_1(\mu)Q_1(\mu)+c_2(\mu) Q_2(\mu)+h.c.)
\nonumber\\
&&Q_1(\mu)=(\bar d'u)_{V-A}(\bar cb)_{V-A}\nonumber\\
&&Q_2(\mu)=(\bar cu)_{V-A}(\bar d'b)_{V-A}\quad.\end{eqnarray}
As an illustrative example only the operators for the decay
process $b\to c\bar u d'$ are exhibited in (\ref{1}).
$d'$ denotes the weak eigenstate of the down-quark. Penguin
operators are neglected. $c_1(\mu)$ and $c_2(\mu)$ are scale-dependent
QCD coefficients known up to next-to-leading order. 
Depending on the process
considered specific forms of the operators can be adopted.
For the process $\bar B^0\to D^+\pi^-$, for instance,
the operator $Q_2$ can be rewritten combining explicitly
the $u$- and $d$-quark fields
\begin{eqnarray}\label{2}
&&Q_2(\mu)=\frac{1}{N_c}Q_1(\mu)+\tilde Q_1(\mu)\nonumber\\
&&\tilde Q_1(\mu)=\frac{1}{2}(\bar d't^au)_{V-A}(\bar c
t^ab)_{V-A}\quad.\end{eqnarray}
$N_c$ denotes the numbers of quark colours and $t^a$ the
Gell-Mann colour SU(3) matrices. The amplitude for the process
$\bar B^0\to D^+\pi^-$ is then
\begin{equation}\label{3}
{\cal A}_{\bar B^0\to D^+\pi^-}=(c_1(\mu)+\frac{1}{N_c}c_2(\mu))
<\pi^-D^+|Q_1|\bar B^0>_\mu+c_2(\mu)
<\pi^-D^+|\tilde Q_1|\bar B^0>_\mu.\end{equation}
For the process $\bar B^0\to D^0\pi^0$, on the other hand,
one has
\begin{equation}\label{4}
{\cal A}_{\bar B^0\to D^0\pi^0}=(c_2(\mu)+\frac{1}{N_c}c_1(\mu))
<\pi^0D^0|Q_2|\bar B^0>_\mu+c_1(\mu)
<\pi^0D^0|\tilde Q_2|\bar B^0>_\mu.\end{equation}
$\tilde Q_2$ consists of the product of colour octet currents
with quark ordering as in $Q_2$. The amplitude for the
process $B^-\to D^0\pi^-$, finally, is simply obtained from
the isospin relation
\begin{equation}\label{5}
{\cal A}_{B^-\to D^0\pi^-}={\cal A}_{\bar B^0\to D^+\pi^-}-\sqrt2
{\cal A}_{\bar B^0\to D^0\pi^0}\quad.\end{equation}

\section{Factorization}

\begin{figure}
\epsfysize=3cm
\centerline{\epsffile{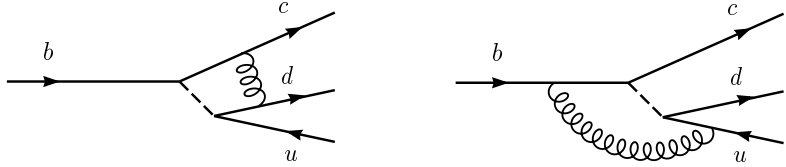}}
\caption{\label{Fig:gluons}
Hard-gluon corrections giving rise to the Wilson coefficients
$c_1(\mu)$ and $c_2(\mu)$ in the effective weak Hamiltonian.}
\end{figure}

How to deal with the complicated matrix elements of four-quark
operators? Because the $(\bar du)$ quark current in $Q_1$
can generate the final $\pi^-$ meson, the matrix element
of $Q_1$ in (3) contains a scale-independent contribution,
the factorization contribution \cite{3}$^-$\cite{5}
\begin{equation}\label{6}
{\cal F}_{(\bar B D)\pi^-}=<\pi^-|(\bar d'u)_{V-A}
|0><D^+|(\bar cb)_{V-A}|\bar B^0>.\end{equation}
Similarly, the matrix element of $Q_2$ in (4) contains the
factorizable part
\begin{equation}\label{7}
{\cal F}_{(\bar B\pi)D}=<D^0|\bar (c u)_{V-A}|0><\pi^0
|(\bar d'b)|\bar B^0>.\end{equation}
Extracting these pieces we can write \cite{6,7}
\begin{eqnarray}\label{8}
&&{\cal A}_{\bar B^0\to D^+\pi}=a_1^{eff}{\cal F}_{(\bar BD)\pi}
\nonumber\\
&&{\cal A}_{\bar B^0\to D^0\pi^0}=a_2^{eff}{\cal F}_{(\bar B\pi)D}
\end{eqnarray}
\begin{eqnarray}\label{9}
&&a_1^{eff}=(c_1(\mu)+\frac{1}{N_c}c_2(\mu))(1+\epsilon_1^{(BD)\pi}
(\mu))+c_2(\mu)\epsilon_8^{(BD)\pi}(\mu)\nonumber\\
&&a_2^{eff}=(c_2(\mu)+\frac{1}{N_c}c_1(\mu))(1+\epsilon_1^{(B\pi)D}
(\mu))+c_1(\mu)\epsilon_8^{(B\pi)D}\quad.\end{eqnarray}
The quantities $1+\epsilon_1(\mu)$ and $\epsilon_8(\mu)$ are the
matrix elements of $Q$ and $\tilde Q$, respectively, divided
by the factorization amplitude. They obey renormalization group
equations \cite{7}. $a_1^{eff}$ and $a_2^{eff}$
are scale-independent, of course. In the large $N_c$ limit neglecting
terms of order $1/N_c^2$, one gets (for a detailed discussion
see Ref. 7)
\begin{eqnarray}\label{10}
a_1^{eff}&=&c_1(\mu)+\zeta(\mu)c_2(\mu)\nonumber\\
a_2^{eff}&=&c_2(\mu)+\zeta(\mu)c_1(\mu)\nonumber\\
\zeta(\mu)&=&\frac{1}{N_c}+\epsilon_8(\mu)\quad.\end{eqnarray}
For $B$-decays putting $\mu=m_b$ one finds to this order
\be\label{11}
a_1^{eff}=1,\quad a_2^{eff}=c_2(m_b)+\zeta(m_b)c_1(m_b)\quad.\ee
$\zeta$ is an unknown dynamical parameter. In general, it will
take different values for
different decay channels. Let us then introduce a process-dependent
factorization scale $\mu_f$ defined by $\epsilon_8(\mu_f)\equiv0$.
The renormali\-zation-group equation for $\epsilon_8(\mu)$ then
gives
\begin{equation}\label{12}
\epsilon_8(\mu)=-\frac{4\alpha_s}{3\pi}\ln\frac{\mu}{\mu_f}+O(\alpha
^2_s)\quad.\end{equation}
For different processes the variation of the factorization scale
$\mu_f$ is expected to scale with 
the change of the energy available for the emitted
quarks. In the 2-body decays under consideration this change
is small compared to the heavy quark mass. Thus, according to (12),
the process dependence of $\epsilon_8(\mu=m_Q)$ is
expected to be very mild, and a single value for $\zeta$ may be
sufficient to describe a number of decays.

On the other hand, $\epsilon_8$ and thus $\zeta$ and $a_2^{eff}$ change
strongly by going from $B$-decays to $D$-decays. In Fig. 2 the ratio
$a_2^{eff}/a_1^{eff}$ is plotted as a function of $\alpha_s(\mu_f)$.
The negative value for this ratio for the case of $D$-decays has
been known for many years \cite{5,8}.
Sign and magnitude imply a small
value of $\zeta$ and lead to a  sizeable destructive amplitude
interference in charged $D$-decays. Since the bulk of $D$-decays
are two-body or quasi two-body decays, this destructive interference
is the main cause for the lifetime difference of $D^+$ and
$D^0$ in full accord with estimates  of the relevant partial
inclusive decay rates \cite{9}.

\begin{figure}
\epsfxsize=7cm
\centerline{\epsffile{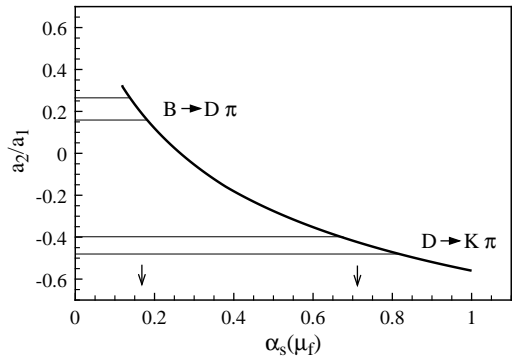}}
\caption{\label{Fig:a2a1}
The ratio $a_2/a_1$ as a function of the running coupling constant
evaluated at the factorization scale. The bands indicate the
phenomenological values of $a_2/a_1$ extracted from $\bar B\to D\pi$
and $D\to K\pi$ decays.}
\end{figure}

For $B$-decays, on the other hand,
the analysis of recent CLEO data \cite{1}
 - to be discussed further below -
indicates a positive value for $a_2^{eff}/a_1^{eff}$ corresponding
to a low $\alpha_s$ value. This is quite in accord with Bjorken's
argument that a fast-moving colour singlet quark pair interacts little
with soft gluons \cite{10}.

From the running of $a^{eff}_2/a_1^{eff}$ in Fig. 2 one can 
also learn that the
ratio of $I=1/2$ to $I=3/2$ decay amplitudes increases with
increasing $\alpha_s (\mu_f)$. In $D$-decays transitions to
$I=1/2$ final states are already much stronger than those to
$I=3/2$ final states. Though non-perturbative QCD prevents an extension
of Fig. 2 to very low energy scales, the trend to $a^{eff}_2/a_1^{eff}
\to-1$, i.e. $c_+(\mu_f)/c_-(\mu_f)\to 0$, is seen. 
Combined with flavour symmetry this limit would give a
complete suppression of $I=3/2$ amplitudes in $D$-decays  
and - for $K$-decays - a strict 
$|\Delta\vec I|=1/2$ selection rule. Indeed, the approximate 
$|\Delta\vec I|=1/2$ rule
observed in strange particle decays found its explanation
in the strong attractive QCD force acting in the 
colour-antitriplet
channel of two quarks \cite{9,11,12}.

\section{Determination of $a_2^{eff}/a_1^{eff}$}

The determination of the ratio $a_2^{eff}/a_1^{eff}$
and in particular of its sign requires measurements of the decays
of the charged heavy meson. In these processes amplitude interference
occurs. The branching ratios normalized to those of the
corresponding neutral meson decays are of the form
\be\label{13}
\frac{\tau(\bar B^0)}{\tau(B^-)}\frac{B(B^-\to D^{(*)0}h^-)}
{B(B^0\to D^{(*)+}h^-)}=1+2x_1\frac{a_2^{eff}}{a_1^{eff}}
+x^2_2\left(\frac{a_2^{eff}}{a_1^{eff}}\right)^2.\ee
Here, $h$ denotes a light meson and $x_1$ and $x_2$ are
process-dependent parameters with $x_1=x_2$ (except for
transitions to two spin-1 particles). For the decay
$B^-\to D^0\pi^-$ one obtains using (\ref{5}), (\ref{8})
\be\label{14}
x_1^{D\pi}=x_2^{D\pi}=\frac{-\sqrt2{\cal F}_{(B\pi)D}}
{{\cal F}_{(BD)\pi}}\quad.\ee
The nominator involves the $\bar B^0\to\pi^0$ current
matrix element and the decay constant of the $D$-meson.
Thus, the calculation of the $x$-values requires model estimates,
in particular, estimates of heavy-to-light form factors.

A detailed analysis of the experimentally determined ratios
of branching ratios (\ref{13}) gives conclusive evidence for
constructive amplitude interference in exclusive $B^-$-decays
\cite{1,7,13}. Moreover, the $x$-values taken from the model of
Ref. 13 leads to a consistent fit of the data with a
positive value for $a_2^{eff}/a_1^{eff}$. Together with the value
of $a_1^{eff}$ obtained from $\bar B^0$-decays the result is
\cite{7}
\be\label{15}
a_1^{eff}|_{Dh}=1.08\pm0.04\qquad a_2^{eff}|_{Dh}=0.23\pm 0.05\quad.\ee
Note that $a_1^{eff}$ was expected to be equal to 1 up to
$1/N_c^2$ corrections. The magnitude of $a_2^{eff}$ is model-dependent
because of the model dependence of the $x$-values. Thus, an
independent determination of these $x$-values is of interest.
One may take advantage of the small ratio $m_c/m_b\simeq 0.2$
for current quark masses and use - for an orientation - the
$c\leftrightarrow d$ exchange symmetry limit. The $B^-$-decay amplitude
consists of a part with the anti-up-quark of
the $B^-$ meson ending up in the charm particle, and of 
a second part where this quark becomes a consituent of the
light meson. The first part is - by replacing the spectator
$\bar u$-quark by a $\bar d$-quark - identical to the corresponding
$\bar B^0$-decay amplitude. The second part, after $c\leftrightarrow
d$ interchange, takes the same form and can also be related to
a $\bar B^0$-decay amplitude. This way one finds
\bea\label{16}
&&x_1=x_2=x,\quad x^{D\pi}=1,\nonumber\\
&&x^{D\rho}=\frac{1}{x^{D^*\pi}}=\left(\frac{B(\bar B^0\to D^{*+}
\pi^-)}{B(\bar B^0\to D^+\rho^-)}\right)^{1/2},\nonumber\\
&&x^{D^*\rho}=1\quad.\eea
The corresponding result for $a_2^{eff}/a_1^{eff}$
is
\be\label{17}
a_2^{eff}/a_1^{eff}|_{Dh}=0.19\pm0.05\quad.\ee
A direct way to obtain $a_2^{eff}$
- but not its sign - is to consider the decays $\bar B^*\to
\bar K^{(*)}J/\psi$ and $\bar B\to \bar K^{(*)}\psi(2S)$.
For these processes the $\gamma$-factors of the outgoing
current generated particles
are much smaller than in the decays considered
before. Taking again the form factors of Ref. 13,
one finds from a fit to the six measured branching ratios
\cite{7}
\be\label{18}
|a_2^{eff}|_{K\psi}=0.21\pm0.01\quad.\ee
The comparison with (\ref{15}) shows that a possible process dependence
of $a_2^{eff}$ cannot be large. There is no evidence for it.
As an independent estimate for the magnitude of $|a_2^{eff}|$ in
$K\psi$ processes one may consider the approximation in which the 
creation operators of the $c$- and $s$- quark fields 
can be interchanged, e.g.
\be\label{19}
|a^{eff}_2/a^{eff}_1|_{K\psi}\approx
\left(\frac{B(\bar B^0\to K^* J/\psi)}{B(\bar B^0\to
D^*D^*_s)}\right)^{1/2}=0.28\pm0.07\quad.\ee
Again we find consistency with the value obtained from
$B\to Dh$ transitions given in (15,17).

\section{Tests of Factorization}

The $B$-meson, because of its large mass, has many decay channels.
Numerous predictions for branching ratios and for the
polarizations of the outgoing particles can be made. All of these
calculations can also serve as tests of factorization. I will
be very short here and have to refer to Ref. 7
for the compilation of branching ratios in tables, for a detailed
discussion and for the comparison with the data. Also discussed in
this reference is the possible influence of final state interactions.
Limits on the relative phases of isospin amplitudes
are given. In
contrast to $D$-decays final state interactions do not seem to
play an essential role in exclusive $B$-decays.

The calculation of non-leptonic transitions which involve
the $B\to D^{(*)}$ form factors presents no problem. These form
factors are well determined \cite{14} using
experimental data and the
heavy quark effective theory \cite{15}. The latter
relates in particular
longitudinal form factors to the transverse ones.

The most direct test of the validity of results
obtained by the factorization method is the one suggested
by Bjorken \cite{10}: the comparison
of the  non-leptonic branching ratio with the corresponding
differential semi-leptonic branching ratio at the relevant
$q^2$ value, i.e.
\be\label{20}
\frac{B(\bar B^0\to D^{(*)+}M^-)}{dB(\bar B^0\to
D^{(*)+}l^-\bar\nu)/dq^2|_{q^2=m^2_M}}=
6\pi^2f^2_M|a^{eff}_1|^2|V_{ij}|^2X^{(*)}_M\quad.\ee
Here $f_M$ is the decay constant of the meson $M$, $V_{ij}$
is an appropriate CKM matrix element, and $X^{(*)}_M$
a number calculable from HQET. $X^{(*)}_M$
is precisely equal to one if $M$ is a vector or axial vector
particle. Within errors Eq. (\ref{20}) is perfectly satisfied by
the data. So far no deviations from
theoretical predictions for $a_1$-type transitions have been
observed.

As mentioned earlier, the predictions for processes which need
heavy-to-light form factors for their evaluation are less reliable.
Nevertheless, the form factors calculated for $q^2$ values
equal to zero in Ref. 5 (BSW-model) and extrapolated
using appropriate pole and dipole formulae \cite{13} give
results still consistent with experiment. They are widely
used. As described in Ref. 7, an
even simpler model which is
free of arbitrary parameters and applicable
to heavy-to-heavy and heavy-to-light transitions also leads to satisfactory
results. It has the advantage of giving very explicit
expressions for the form factors and has a wide field of
applications. These models are intended to give an overall
picture. They should be complemented by detailed
investigations of important special processes using more
sophisticated methods.

Non-leptonic decays to two spin-1 particles need a
separate discussion. Here one has 3 invariant amplitudes corresponding
to outgoing $S$, $P$, and $D$-waves. Factorization fixes the
strength of all 3 amplitudes and thus predicts besides the
decay rates also the polarization of the
outgoing particles. In factorization approximation the
polarization is the same as the one occurring in
the corresponding semi-leptonic decay at the appropriate
$q^2$ value. For $B\to D^*V$ decays the theoretical predictions
have very small errors only.
For instance, the transverse polarization
of the final particles in the decay $\bar B^0\to D^*\bar D
_s^*$ is $(48\pm1)$ \%, if factorization holds \cite{7}.
These decays allow, therefore,
subtle tests of factorization.

The polarization of the final state particles is very sensitive to
non-factorize\-able contributions and final state interactions.
Non-factorizeable contributions to the amplitude will, in general,
have a different structure for different partial waves compared
to the factorizeable contributions. Likewise,
final state interactions are different for different
partial waves. They may change $S$- into $D$-waves even without
changing the total decay rate. A case of particular interest is
the polarization of the $J/\psi$ particle in the decay $B\to K^* J/\psi$.
Most models predict a longitudinal polarization
of around 40 \% (the ones mentioned above predict 35 \% and 48 \%,
respectively). The world average, on the other hand, is
$(78\pm7)$ \%. As reported at this conference \cite{16}, a recent CLEO
measurement \cite{} gives $(52\pm7\pm4)$ \%, however. Although a
clear picture has not yet emerged, it is possible that the first
deviations from factorization predictions in $B$-decays will be seen
in polarization data.

\section{Use of Factorization}

As we have seen, the calculation of two-body and quasi
two-body $B$-decays by the factorization method gave results
in remarkable agreement with experiment. Within present experimental
error limits no deviations have been found -- with the
possible exception of the $J/\psi$ polarization mentioned
above. Besides providing many predictions for not yet measured
decays, factorization can, therefore, also be used
to determine unknown decay constants. A case in point 
is the determination of
the decay constants of the $D_s$ and $D_s^*$ particles.
One may use again Eq. (\ref{20}) with $a_1^{eff}$ taken from
$B\to Dh$ decays (Eq. (\ref{15})) \cite{17}. But an even more precise
determination is obtained if we consider ratios of
non-leptonic decay rates, comparing decays to $D_s, D^*_s$ with
those to light mesons. Here $a_1^{eff}$ cancels and, presumably,
also some of the experimental systematic errors. One finds \cite{7}
\be\label{21}
f_{D_s}=(234\pm25)\ {\rm MeV}, \quad f_{D^*_s}=
(271\pm33)\ {\rm MeV}\quad.\ee
The value for $f_{D_s}$ is in excellent agreement with the value
$f_{D_s}=(241\pm37)\ {\rm MeV}$ obtained from the
leptonic decay of the $D_s$ meson \cite{18}.

There are several other decay constants which can be determined
this way. Of particular interest are the decay constants of
$P$-wave mesons like the $a_0,\ a_1,\ K^*_0,\ K_1$ particles.
The decay constant of the pseudovector meson $a_1$ can be
obtained from the ratio $B(\bar B^0\to D^{*+}a^-_1)
/B(\bar B^0\to D^{*+}\rho^-)$.
The result
\be\label{22}
f_{a_1}=(256\pm40)\ {\rm MeV}\ee
is in agreement with the large value $f_{a_1}=(229\pm10)\ {\rm MeV}$
obtained from $\tau$-decay.

Another possible application of factorization is the determination
of form factors not accessible in semi-leptonic decays.
For example, one can obtain the ratio of a neutral current form factor
at two different $q^2$ values.

\section{Summary}

The matrix elements of non-leptonic exclusive decays are
notoriously difficult to calculate. The concept of factorization,
however, provides a connection to better known objects. 
At least for energetic $B$-decays factorization is found to be extremely
useful, if properly applied and interpreted. It passed
many tests. Thus, it allows reliable predictions for many
decay channels as well as the determination of decay
constants which are difficult to measure otherwise. 
Transitions to two vector particles need special attention. They
are more sensitive
to non-factorizeable contributions and final state interactions.

The parameter $a_1^{eff}$ is predicted to be $1$
apart from $1/N^2_c$ corrections and to be practically
process-independent. Experiments support these expectations. The
particularly interesting parameter $a_2^{eff}$, within
errors, does also not show a process dependence. A good
test, however, would require a better knowledge of heavy-to-light
form factors.
The positive value of $a_2^{eff}/a_1^{eff}$ found to describe
exclusive $B$-decays is most remarkable. The obvious interpretation
is that a fast-moving colour singlet quark pair interacts
little with soft gluons. Thus, for these transitions, the parameter
 $\zeta$ is
close to the naive factorization prediction $\zeta=1/3$.
The constructive interference observed in energetic two-body $B^-$-decays
does not imply that the lifetime of the $B^-$ meson should be
shorter than the lifetime of the $\bar B^0$ meson: The majority
of transitions proceed into multi-body final states. For these
the relevant scale may be lower than $m_b$ leading to destructive
interference. Moreover, there are many decay channels for which
interference cannot occur.
The change of the constants $a_1^{eff},\ a_2^{eff}$ with the
available kinetic energy of the outgoing quarks (or the effective
$\alpha_s$) is very intriguing. In $D^+$-decays one has a
sizeable destructive amplitude interference which  
causes the lifetime difference of $D^0$ and $D^+$, and
$I=1/2$ amplitudes to be large
compared to $I=3/2$ amplitudes. In strange particle decays
the observed dominance of $|\Delta\vec I|=1/2$ transitions 
is the most spectacular 
manifestation of these QCD effects.
One sees a unified picture of exclusive non-leptonic decays
which ranges from very low scales to the large energy scales
relevant for $B$-decays.
\section*{Acknowledgement}
The work reported here was performed in a most pleasant
collaboration with Matthias Neubert. The author also likes
to thank Tom Browder and Sandip Pakvasa for the invitation
to this stimulating meeting.

%\begin{figure}
%\rule{5cm}{0.2mm}\hfill\rule{5cm}{0.2mm}
%\vskip 2.5cm
%\rule{5cm}{0.2mm}\hfill\rule{5cm}{0.2mm}
%\psfig{figure=filename.ps,height=1.5in}
%\caption{Radiative (off-shell, off-page and out-to-lunch) SUSY Higglets.
%\label{fig:radish}}
%\end{figure}

%\section*{Acknowledgments}

%\section*{Appendix}
% We can insert an appendix here and place equations so that they are
%given numbers such as Eq.~\ref{eq:app}.
%\be
%x = y.
%\label{eq:app}
%\ee

%\bibitem{bu}J.D. Bjorken and I. Dunietz, \Journal{\PRD}{36}{2109}{1987}.
%\bibitem{bd}C.D. Buchanan {\it et al}, \Journal{\PRD}{45}{4088}{1992}.
%\end{thebibliography}
\end{document}